# Quasiparticle scattering from topological crystalline insulator SnTe (001) surface states

Duming Zhang,[1,2] Hongwoo Baek,[1,3] Jeonghoon Ha,[1,2] Tong Zhang,[1,2,*] Jonathan E. Wyrick,[1] Albert V. Davydov,[4] Young Kuk,[3] and Joseph A. Stroscio[1,†]

[1]Center for Nanoscale Science and Technology, National Institute of Standards and Technology, Gaithersburg, MD 20899, USA
[2]Maryland NanoCenter, University of Maryland, College Park, MD 20742, USA
[3]Department of Physics and Astronomy, Seoul National University, Seoul, 151-747, Korea
[4]Material Measurement Laboratory, National Institute of Standards and Technology, Gaithersburg, MD 20899, USA

Recently, the topological classification of electronic states has been extended to a new class of matter known as topological crystalline insulators. Similar to topological insulators, topological crystalline insulators also have spin-momentum locked surface states; but they only exist on specific crystal planes that are protected by crystal reflection symmetry. Here, we report an ultra-low temperature scanning tunneling microscopy and spectroscopy study on topological crystalline insulator SnTe nanoplates grown by molecular beam epitaxy. We observed quasiparticle interference patterns on the SnTe (001) surface that can be interpreted in terms of electron scattering from the four Fermi pockets of the topological crystalline insulator surface states in the first surface Brillouin zone. A quantitative analysis of the energy dispersion of the quasiparticle interference intensity shows two high energy features related to the crossing point beyond the Lifshitz transition when the two neighboring low energy surface bands near the $\bar{\mathrm{X}}$ point merge. A comparison between the experimental and computed quasiparticle interference patterns reveals possible spin texture of the surface states.

[*] Current address: Department of Physics, Fudan University, Shanghai 200433, China.
[†] To whom correspondence should be addressed: joseph.stroscio@nist.gov

## I. Introduction

Topological insulators are a new classification of matter characterized by a bulk insulating gap and gapless surface states protected by time reversal symmetry.[1–3] This is realized by spin-orbit coupling induced band inversion with an odd number of Dirac cones. Recently, the topological classification of materials has been extended to a new phase of matter, topological crystalline insulators.[4,5] In contrast to topological insulators, topological crystalline insulators arise from crystal reflection symmetry and are characterized by topological surface states with an even number of Dirac cones. The first topological crystalline insulator was predicted in the SnTe class of materials,[5] the surface states of which were soon observed in $Pb_{1-x}Sn_xSe$,[6] SnTe,[7] and $Pb_{1-x}Sn_xTe$[8,9] bulk crystals by angle-resolved photoemission spectroscopy (ARPES). The surface states of this class of topological crystalline insulators have also been further studied by scanning tunneling microscopy (STM)[10–12] as well as electrical transport.[13,14] Most of the studies so far have focused on cleaved bulk samples. However, there are several advantages to grow these topological materials in the form of nanostructures and thin films. First, by going to lower dimensions, the surface contribution can potentially be enhanced with increased surface-to-volume ratio.[14,15] Furthermore, compared to bulk crystals, it is much easier to fabricate devices made from nanostructures and to interface them with other materials such as superconductors[16,17] and magnetic materials.[18] Motivated by these advantages, we performed an STM study of the topological surface states of SnTe nanoplates.

Here, we report synthesis and *in-situ* STM measurements on single crystalline SnTe nanoplates synthesized by molecular beam epitaxy (MBE). We carried out Fourier transform scanning tunneling spectroscopy (FT-STS) using an ultra-low temperature STM on the SnTe (001) surface. We observed quasiparticle interference patterns in the differential tunneling conductance, *dI/dV*, maps, which can be interpreted in terms of scattering among the four Fermi pockets of the topological surfaces states in the first surface Brillouin zone of (001) planes. A quantitative analysis of the energy dispersion of the quasiparticle interference intensity reveals features associated with the high energy crossing point in the surface bands.

## II. Synthesis and Characterization of SnTe Nanoplates

The SnTe nanoplates studied in this work were synthesized by MBE using a high purity SnTe compound source (pieces, 99.999%) and a separate elemental Sn source (shots, 99.999%). After a 30-minute anneal of a graphitized 6H-SiC (0001) substrate at 300 °C, SnTe was deposited onto the substrate at ≈ 230 °C with a typical growth rate between 0.3 nm/min and 0.7 nm/min and a Sn to SnTe flux ratio in the range of 0 % to 15 %. We note that the compensation of extra Sn flux (up to 15 % in flux ratio) did not seem to reduce the overall *p*-doping concentration in the samples. SnTe has a cubic rock salt crystal structure with a lattice constant $a = 0.633$ nm. A schematic of the SnTe (001) surface is illustrated in Fig. 1(a). Reflection high energy electron diffraction (RHEED) was used to monitor the growth front, which indicates single crystal growth mode [Fig. 1(b)]. *Ex-situ* atomic force microscopy (AFM) [Fig. 1(c)] shows that all these nanoplates are roughly square shaped, suggesting they have a preferential out-of-plane orientation along the <001> direction. This preferential orientation of the nanoplates is confirmed by X-ray diffraction (XRD) measurement, showing only the {001} reflections [Fig. 1(d)]. Electron backscatter diffraction measurements on individual nanoplates also confirmed the nanoplates were single crystalline with (001) oriented top surfaces.

### III. Experimental Results

#### A. Sn vacancy defects

After the growth of SnTe nanoplates, the sample was immediately transferred from the MBE chamber to an interconnected 10 mK ultra-low temperature STM without breaking vacuum.[19] The STM topographic image in Fig. 2(a) reveals the SnTe (001) surface with large atomically flat terraces separated by a step height of one half unit cell (≈ 0.32 nm, see Fig. 1(a)). Fig. 2(b) shows a smaller scale STM topographic image of the SnTe (001) surface, with a number of defects, which is a characteristic of these samples. At positive sample bias ($V_{bias} = 0.8$ V), only the Sn sublattice is mainly revealed as electrons tunnel into the empty states of the sample. The lattice spacing is consistent with (110) interplanar distance $d = a/\sqrt{2} = 0.45$ nm as illustrated in Fig. 1(a). Apart from some adatoms at the surface, many vacancies, presumably Sn vacancies, are clearly visible. This is consistent with as-grown SnTe bulk crystals, which are typically *p*-doped due to Sn vacancies.[7] The long wavelength roughness at the SnTe surface (root mean square roughness = 8.8 pm) may be due to the underlying structure of multi-layer graphene grown on the SiC substrate, which was used as a substrate for the SnTe growth.

The (001) surface of SnTe has been predicted[5] and shown[7] to have topological surface states, a schematic of which is shown in Fig. 2(c). By rotation symmetry, there are four low energy bands near the four equivalent $\bar{X}$ points in the first surface Brillouin zone indicated by the shaded square. Due to crystal reflection symmetry, there are two low energy surface bands located symmetrically about the $\bar{X}$ point along the $\bar{\Gamma}\bar{X}$ direction. The positions of the neighboring zero energy crossing points are denoted as $\bar{\Lambda}_1$ and $\bar{\Lambda}_2$. When the energy is increased/decreased from the zero energy crossing point, the two low energy surface bands on both sides of the $\bar{X}$ point touch each other, exhibiting a Lifshitz transition, where two electron/hole pockets reconnect to form a large electron/hole and a small hole/electron pocket centered at the $\bar{\mathrm{X}}$ point. When the energy is further increased/decreased from the zero energy crossing point, the high energy surface bands start from the high energy crossing points at the $\bar{\mathrm{X}}$ point.

The local density of states of the SnTe nanoplates was obtained by measuring the bias dependent *dI/dV* spectra with a lock-in amplifier. Fig. 2(d) is a typical single point spectrum with a minimum at ≈ 350 meV. The inset shows a simplified schematic of the two low energy surface bands located symmetrically about the $\bar{\mathrm{X}}$ point along the $\bar{\Gamma}\bar{\mathrm{X}}$ direction. Comparing with this surface state band diagram, we assign this *dI/dV* minimum as the zero energy crossing point, $E_0$. The position of the zero energy crossing point indicates that our nanoplates are *p*-doped, consistent with the observation of large amount of Sn vacancies at the surface. Similar to the STM work on cleaved bulk $Pb_{1-x}Sn_xSe$,[10] we did not observe strong features in the spectra associated with the Lifshitz transition. However, features related to the Lifshitz transition were recently observed in Ref. [11] in $Pb_{1-x}Sn_xSe$, and it is uncertain why they are not visible in our spectra on SnTe. Evidence for the Lifshitz transition is seen in our data in the analysis of the quasiparticle interference presented in section III, below. Based on the position of the zero energy crossing point and the band gap (≈ 0.18 eV) of bulk SnTe, the kink near the zero bias in the spectrum is possibly related to states in the bulk valence band.

The *p*-doping is generally attributed to the tendency to grow nonstoichiometric SnTe with Sn vacancies, which can be observed in the STM images. Figures 3(a) and 3(c) show a series of high-resolution topographic images of atomic defects from the same area at different sample biases. At high positive sample biases [Fig. 3(a)], electrons tunnel from the tip into the empty states of the conduction band and image mainly the Sn sublattice. The absent rows of Te atoms are between the Sn rows as indicated by the white dashed lines in Fig. 3(a). We can clearly see two missing Sn atoms at the surface indicated by the red arrows, which we identify as Sn vacancies at the surface layer. This can be further confirmed by imaging the same area at negative sample biases [Fig. 3(c)]. At a negative bias, electrons tunnel from the filled states of the valence band into the tip and the Te sublattice is enhanced in the topographic images. By comparing the position of the two sublattices, we can identify that the Sn vacancies at the top surface layer are located at the center of four neighboring Te atoms [Fig. 3(c)]. The Te atom rows are indicated by the white dashed lines in Fig. 3(c). We note that the Sn sublattice in the image has already switched to the Te sublattice with the Sn vacancy sites at the center of four neighboring Te atoms for energies near the zero energy crossing point. This is likely due to hybridization effects induced by the large spin-orbit coupling across the bulk band gap.

To further verify the Sn vacancies observed at the SnTe (001) surface, we simulated the STM images in Figs. 3(b) and 3(d) as integrated charge density iso-surfaces from density function theory (DFT) calculations using the generalized gradient approximation for the exchange-correlation functional.[20–24] The atomic configuration consisted of a 3 layer slab with 3×3 surface unit cells having one surface Sn atom removed while all other atoms were fixed at their bulk positions. A real space grid spacing of 30 pm and a *k*-point mesh of 5×5×1 were used with a super-cell that included 1 nm of vacuum along the direction normal to the surface. For comparison to the experimentally measured STM images, each simulated image has its super-cell repeated 3 times in each direction (for a total of 9×9 unit cells with 9 vacancies per image). As we can see in Figs. 3(b) and 3(d), the simulated STM images correctly capture the main features of the experimental data: vacancies at Sn atomic sites at positive sample biases occur in line with the Sn rows and in between the Te atomic sites at negative sample biases. There are also several other types of defects indicated by the green arrows in Fig. 3(a), which could be anti-sites or vacancies underneath the surface layer. Identification of these different types of defects would require further DFT simulations.

### B. Quasiparticle interference

The presence of these surface defects is actually useful for studying the Fermi surface of SnTe through electron scattering. Indeed, clues of such scattering from defects can already be seen in topographic images as interference patterns around the Sn vacancies [Fig. 3(a)]. To better understand the scattering process, we study quasiparticle interference patterns obtained from FT-STS maps, which can provide the real space and momentum space electronic structure information simultaneously. This method has been applied to study noble metal surface states,[25] high-$T$c superconductors,[26] graphene[27,28] as well as topological materials.[29] Figure 4 shows interference patterns in STS ($dI/dV$) maps at different sample biases with respect to the zero energy crossing point $E_0$ of 350 meV. The image size (47.5 nm $\times$ 47.5 nm) and resolution (475 pixels $\times$ 475 pixels) were chosen to cover at least the first two Brillouin zones with a resolution better than 1 % of the Brillouin zone size. By taking the Fourier transform of the STS maps, we can measure the quasiparticle scattering vectors as a difference of the initial and final wave vectors, $\mathbf{q} = \mathbf{k}_f - \mathbf{k}_i$, for elastic scattering. Combined with information on the surface band structure in $k$ space, we can study the Fermi surface and possible spin textures of the surface states. Figure 5(a) shows a 3D illustration of the low energy surface bands below the Lifshitz transition in the first surface Brillouin zone. Theoretically expected spin texture of the topological crystalline insulator surface states is indicated by the small arrows. Possible elastic scattering between $\mathbf{k}_j(E)$ is indicated by the color coded arrows: $q_1$ represents intra-cone scattering and $q_2 - q_4$ represent inter-cone scattering. In Fig. 5(b), we show the schematic contours of constant energy of the surface states near the zero energy crossing point enclosed in the first surface Brillouin zone with size of $2\pi/d \times 2\pi/d$. We also include $q'_3$ which represents the inter-cone scattering between the two neighboring low energy bands located symmetrically about the $\bar{X}$ point. Due to the periodic Brillouin zone boundary condition, $q'_3$ is equivalent to $q_3$. The quasiparticle interference patterns from the first surface Brillouin zone can be described in terms of the $q$ wave vectors within a box with size of $4\pi/d \times 4\pi/d$ [Fig. 5(c)]. The intra-cone scattering ($q_1$) is represented by the circle symbol at the center of the box and the inter-cone scattering ($q_2 - q_4$) is represented by the square and star symbols along the four equivalent directions of $\bar{\Gamma}\bar{X}$ and $\bar{\Gamma}\bar{M}$, respectively. $q'_3$ is also expected to be along the $\bar{\Gamma}\bar{X}$ direction but closer to $q_1$, as indicated by the dashed star symbols. By rotation and crystal reflection symmetry, there are only two sets of different inter-cone scattering wave vectors, $q_2$ and $q_3$ ($q'_3$). As the energy is moved away from the zero energy crossing point, the general trend of the quasiparticle interference pattern is that the disk size of the $q$ vectors increases as the Fermi pockets become larger while the center position of the $q$ vector disks remains relatively unchanged.

Figure 6 shows raw unfiltered quasiparticle interference patterns obtained by Fourier transforming the $dI/dV$ maps in Fig. 4. The four bright spots near $q_3$ and the four bright spots in the middle of the image edges are the Bragg peaks originating from the atomic corrugation of the underlying SnTe lattice [see arrows in Fig. 6(a)]. The intra-cone scattering $q_1$ wave vector is located at the center, which is accompanied by intensity from long wavelength modulations from disorder in the sample. The two pairs of equivalent inter-cone scattering $\pm q_2$ and $\pm q_3$ wave vectors are also observed at the expected positions. The elongated shape of $q_2$ and $q_3$ indicates the anisotropy of the surface bands. Although $q'_3$ is also expected according to the schematic in Fig.

5, it was not observed. This may be related to matrix element effects resulting in a different amplitude of $q'_3$. As the energy is stepped through the zero energy crossing point, the intensity of the quasiparticle interference patterns decreases first and then increases, with a minimum around the zero energy crossing point. We also note an intensity asymmetry in the quasiparticle interference patterns with respect to $q_2$ and $q_4$, which occurs near the zero energy crossing point [Figs. 6(b) - 6(d)]. The origin of this asymmetry is unknown at present, but may be due to tip asymmetries, or rhombohedral distortion in the SnTe atomic lattice.

To further study the Fermi surface of the SnTe (001) topological surface states, we plot the energy dispersion of the quasiparticle interference pattern intensity along the $\bar{\Gamma}\bar{M}$ ($q_2$) and $\bar{\Gamma}\bar{X}$ ($q_3$) directions in Figs. 7(a) and 7(b), respectively. First, let's take a look at the $q_1$ feature near zero wave vector transfer, $q = 0$. Near the zero energy crossing point, the $q_1$ intensity for both directions in Figs. 7(a) and 7(b) is dominated by intensity due to long wavelength modulations from disorder in the sample. However, as the energy is increased/decreased away from the zero energy crossing point, the $q_1$ feature starts to disperse as the Fermi pockets of the surface states grow larger, and is particularly clear in Fig. 7(b). In Fig. 7(a), we can clearly see the $\pm q_2$ feature at $\approx \pm 0.75$ Å$^{-1}$. The intensity of $q_2$ is weaker and its peak width is narrower when the energy is close to the zero energy crossing point. Below the zero energy crossing point, the $q_2$ peak width increases as the Fermi pockets become larger while its position remains almost unchanged as expected. For dispersion along the $q_3$ direction [Fig. 7(b)], the high intensity features at $\pm 1.45$ Å$^{-1}$ are the Bragg peaks. Features of $\pm q_3$ wave vectors are located close to the Bragg peaks at $\approx \pm 1.09$ Å$^{-1}$, which do not disperse much as the energy is changed. We find the separation between the zero energy crossing point $\bar{\Lambda}_1$ and the $\bar{X}$ point in $k$ space to be $0.180 \pm 0.003$ Å$^{-1}$,[30] which is slightly larger than the value of $0.15 \pm 0.01$ Å$^{-1}$ obtained by ARPES from SnTe bulk crystal.[9] The larger $\bar{\Lambda}_1\bar{X}$ distance in our sample is likely due to the higher *p*-doping concentration as Ref. [9] shows that less *p*-doped samples tend to have smaller $\bar{\Lambda}_1\bar{X}$ distance. At $E - E_0 \leq -175$ meV, there are two distinct features near $q_3$ and $q_1$ dispersing in opposite $q$ directions as a function of energy. As we will discuss below and in Figs. 8 and 9, these two features, noted as $q_{H-}$ and $q'_{H-}$ respectively, are related to the high energy crossing point beyond the Lifshitz transition energy.

From the energy dispersion of quasiparticle interference pattern intensity plots, we can in principal extract information about the Fermi velocities along the high symmetry ($q_2$ and $q_3$) directions. However, due to the strong domination of intensity due to disorder near $q = 0$ at energies close to the zero energy crossing point, it is not reliable to deduce the Fermi velocities from the $q_1$ wave vector. Instead, we show that the Fermi velocity along the $\bar{\Gamma}\bar{X}$ and $\bar{X}\bar{M}$ directions can be possibly deduced from the $q_{H-}$ and $q'_{H-}$ features at energies far away from the zero energy crossing point observed in Fig. 7(b). Fig. 8(a) shows line cuts along the $\bar{\Gamma}\bar{X}$ ($q_3$) direction in the energy range from $E - E_0 = -300$ meV to $E - E_0 = -200$ meV. The peaks near the Bragg peak and $q = 0$ are denoted as $q_{H-}$ and $q'_{H-}$, respectively, as indicated by the arrows in Fig. 8(a). As we can see clearly from Figs. 7(b) and 8(a), these two features disperse in opposite $q$ directions. The inset of Fig. 8(a) plots the sum of $q_{H-}$ and $q'_{H-}$ peak positions versus energy, which is very close to the Bragg peak position (red dashed line), i.e., $2\pi/d$. Figs. 8(b) and 8(c) plot energy dispersion of the $q_{H-}$ and $q'_{H-}$ peak positions. Linear fits to the data yield the same slope within error but with opposite signs.

These results suggest that the $q_{H-}$ and $q'_{H-}$ features originate from the same scattering mechanism. By examining possible scatterings along the $\overline{\Gamma X}$ direction above the Lifshitz transition, we discuss below two possible interpretations of the data in terms of relevant critical spanning vectors and show these two features can be related to the high energy crossing point formed when the two neighboring low energy surface bands ($\overline{\Lambda}_1$ and $\overline{\Lambda}_2$) near the $\overline{X}$ point merge together. Figure 9(a) shows schematic contours of constant energy of the surface bands above the high energy crossing point in *k* space. The box indicates the first surface Brillouin zone. Scattering along the $\overline{\Gamma X}$ direction is dominated by the critical spanning vectors along line cuts of $\overline{\Gamma X}$ (blue dashed line) and $\overline{XM}$ (red dashed line). In the following paragraphs, we will discuss two possible interpretations for the $q_{H-}$ and $q'_{H-}$ features: I. Scattering dominated by the critical spanning vectors along the $\overline{\Gamma X}$ line cut; and II. Scattering dominated by the critical spanning vectors along the $\overline{XM}$ line cut.

### B - I. Scattering along the $\overline{\Gamma X}$ line cut

First we focus on the $\overline{\Gamma X}$ line cut in Fig. 9(b). There are two linear surface bands offset vertically by $2E_{H+}$ in energy at the $\overline{X}$ point. We refer to the cone (pocket from branches 2 and 3) with the crossing point at $E_{H+}$ as $CONE_{H+}$ and the cone (pocket from branches 1 and 4) with the crossing point at $E_{H-}$ as $CONE_{H-}$ (Unless noted elsewhere, +/- denotes features above/below the zero energy crossing point). Due to the periodic Brillouin zone boundary conditions, possible scatterings among the surface bands on the opposite zone boundaries can be reduced to intra-cone scatterings of $CONE_{H+}$ ($q_{H+}^{\overline{\Gamma X}}$, solid green arrows) and $CONE_{H-}$ ($q_{H-}^{\overline{\Gamma X}}$, solid green arrows). We note $q_{H\pm}^{\bar{\Gamma}\bar{X}}$ at $E_{H-} < E < E_{H+}$ is a subset of $q_3$ defined earlier in Fig. 5. The $q_1^{\bar{\Gamma}\bar{X}}$ wave vector is indicated by the solid red arrow.

After identifying the possible scattering along the $\bar{\Gamma}\bar{X}$ line cut, we can translate this from $k$ space to $q$ space to understand the $q_{H-}$ and $q'_{H-}$ features observed in Fig. 7(b). Figure 9(d) shows the energy dispersions of the critical spanning vectors along the $\bar{\Gamma}\bar{X}$ ($q_3$) direction using energy-momentum dispersions of the surface bands from Ref. [31]

$$E_{H,L}(k) = \sqrt{m^2 + \delta^2 + v_x^2 k_x^2 + v_y^2 k_y^2 \pm 2\sqrt{m^2 v_x^2 k_x^2 + (m^2 + \delta^2) v_y^2 k_y^2}}, \quad (1)$$

with the parameters[31] $m = -70$ meV, $\delta = 26$ meV, $v_x = 2.40$ eV·Å, and $v_y = 1.40$ eV·Å. Solid lines represent the energy dispersions of the critical spanning vectors along the $\bar{\Gamma}\bar{X}$ line cut with $k_x = 0$ in Eq. (1). All the features from the $\bar{\Gamma}\bar{X}$ line cut have linear energy dispersion with the same slope of $v_y/2$. $q_3^{\bar{\Gamma}\bar{X}}$ (solid green lines) is located 0.1 Å$^{-1}$ away from $q = 0$ and $2\pi/d$ at the zero energy crossing point. As energy is increased/decreased from the zero energy crossing point, one branch of $q_3^{\bar{\Gamma}\bar{X}}$ extends out to $q = 0$ or $2\pi/d$ at $E_{H\pm} = \pm 75$ meV and then folds back as $q_{H\pm}^{\bar{\Gamma}\bar{X}}$ with the same slope. $q_1^{\bar{\Gamma}\bar{X}}$ (solid red lines) originates from $q = 0$ and $2\pi/d$ at the zero energy crossing point and disperse as a function of energy with the same slope as $q_3^{\bar{\Gamma}\bar{X}}$.

With the energy dispersions of the critical spanning vectors along the $\bar{\Gamma}\bar{X}$ line cut described above, we next aim to identify the origin of the $q_{H-}$ and $q'_{H-}$ features. As discussed in the previous paragraph, the slope of all the features is $v_y/2$. Therefore, we can obtain the Fermi velocities associated with the $q_{H-}$ and $q'_{H-}$ peaks from the linear fits in Figs. 8(b) and 8(c):

$v_{\text{F-H}} = 1.51 \pm 0.08$ eV·Å and $v'_{\text{F-H}} = 1.28 \pm 0.28$ eV·Å.[32] We note that $q_{\text{H+}}$ and $q'_{\text{H+}}$ are also visible at $E - E_0 > 75$ meV in Fig. 7(b). However, we were not able to deduce reliable Fermi velocities due to limited data range. Depending on the choice of parameters for Eq. (1), there can be multiple energy dispersions of different critical spanning vectors close to the observed $q_{\text{H-}}$ and $q'_{\text{H-}}$ peaks. Although it may be difficult to distinguish them in the high energy range, it is possible to identify these features by examining their intercepts to $q = 0$ and $2\pi/d$. As shown in Fig. 9(d), The intercepts at $q = 0$ or $2\pi/d$ for $q_{\text{H-}}^{\overline{\Gamma X}}$, $q_1^{\overline{\Gamma X}}$, and $q_{\text{H+}}^{\overline{\Gamma X}}$ are $E_{\text{H-}}$ (negative value), 0, and $E_{\text{H+}}$ (positive value), respectively. The linear fits in Figs. 8(b) and 8(c) yield the intercept of $q_{\text{H-}}$ to the Bragg peak as $E_{\text{H-}} = -71 \pm 9$ meV and the intercept of $q'_{\text{H-}}$ to $q = 0$ as $E'_{\text{H-}} = -97 \pm 36$ meV, respectively.[32] The negative intercepts of the $q_{\text{H-}}$ and $q'_{\text{H-}}$ peaks rule out $q_1^{\overline{\Gamma X}}$ and $q_{\text{H+}}^{\overline{\Gamma X}}$. Therefore, we can attribute the experimental $q_{\text{H-}}$ and $q'_{\text{H-}}$ peaks to the $q_{\text{H-}}^{\overline{\Gamma X}}$ critical spanning vector for this case of $\overline{\Gamma X}$ scattering. This choice gives the Fermi velocity along the $\overline{\Gamma X}$ direction determined from $q_{\text{H-}}$ as $v_y = 1.51 \pm 0.08$ eV·Å and the high energy crossing point $E_{\text{H-}} = -71 \pm 9$ meV.[32] The Fermi velocity deduced from this case is close to $v_y = 1.3$ eV·Å suggested in Ref. [31] and $v_y = 1.1 \pm 0.3$ eV·Å obtained from $\text{Pb}_{0.6}\text{Sn}_{0.4}\text{Te}$ bulk crystals[8] by ARPES measurements, but smaller than $v_y = 2.5 \pm 0.3$ eV·Å obtained from SnTe bulk crystals.[7] To compare directly with the experimental energy dispersion of the quasiparticle interference intensity, we plot the energy dispersion of the critical spanning vectors for the $\overline{\Gamma X}$ line cut on top of the experimental data for the range of $q < 0$ (left hand portion) in Fig. 9(e) with $v_y = 1.51$ eV·Å and $\sqrt{m^2 + \delta^2} = 71$ meV. The $q_{\text{H-}}$ and $q'_{\text{H-}}$ features are well described by the energy dispersion of the $q_{\text{H-}}^{\overline{\Gamma X}}$ critical spanning vector (solid green lines). However, using relations $\bar{\Lambda}_{1,2} = \left(0, \pm\sqrt{m^2+\delta^2}/v_y\right)$ and $E_{H\pm} = \pm\sqrt{m^2+\delta^2}$ from Ref. [31], we get $\bar{\Lambda}_1\bar{X} = 0.047 \pm 0.006$ Å$^{-1}$.[33] This value is much smaller than the value of $0.180 \pm 0.003$ Å$^{-1}$ obtained from the quasiparticle interference patterns in Fig. 6 and energy dispersion in Fig. 7(b). As indicated in Fig. 9(e) by the arrow, the expected $q_3$ location at the zero energy crossing point is closer to the Bragg peak than the experimental result.

### B - II. Scattering along by the $\overline{XM}$ line cut

We next explore the possibility of scattering along the $\overline{XM}$ line cut. Figure 9(c) shows the schematic critical spanning vectors along the $\overline{XM}$ line cut. Possible scatterings are indicated by the color coded arrows: $q_{H+}^{\bar{X}\bar{M}}$ represents scattering between branches 2 and 3 above $E_{H+}$; $q_{1L+}^{\bar{X}\bar{M}}$ represents scattering between branches 1 and 3 or branches 2 and 4; $q_{2L+}^{\bar{X}\bar{M}}$ represents scattering between branches 1 and 4; and $q_{3L+}^{\bar{X}\bar{M}}$ represents scattering between branches 3 and 2 at $E_{L+} < E < E_{H+}$. We note that for $E_{L-} < E < E_{L+}$, the scattering along the $\overline{XM}$ direction for the two separate cones located symmetrically about the $\bar{X}$ point is essentially a subset of $q_1$.

Using Eq. (1) with $k_y = 0$ and parameters described above, we plot the energy dispersion of the critical spanning vectors along the $\overline{XM}$ line cut as dashed lines in Fig. 9(d). The $q_{H\pm}^{\bar{X}\bar{M}}$ (dashed green lines) originates from $q = 0$ and $2\pi/d$ at $E_{H\pm} = \pm 75$ meV while $q_{1L\pm}^{\bar{X}\bar{M}}$ (dashed red lines) originates from $q = 0$ and $2\pi/d$ at $E_{L\pm} = \pm 26$ meV. Both $q_{2L\pm}^{\bar{X}\bar{M}}$ and $q_{3L\pm}^{\bar{X}\bar{M}}$ originate at 0.058 Å$^{-1}$ away from $q = 0$ and $2\pi/d$ at $E_{L\pm} = \pm 26$ meV. As the energy is moved away from the zero energy crossing point, $q_{3L\pm}^{\bar{X}\bar{M}}$ extends to $q = 0$ and $2\pi/d$ at $E_{H\pm} = \pm 75$ meV while $q_{1L\pm}^{\bar{X}\bar{M}}$ and $q_{2L\pm}^{\bar{X}\bar{M}}$ features disperse with the same curvature as $q_{H\pm}^{\bar{X}\bar{M}}$. When energy is far away from the zero energy crossing point ($E - E_0 \gg \delta$), $q_{H\pm}^{\bar{X}\bar{M}}$, $q_{1L\pm}^{\bar{X}\bar{M}}$, and $q_{2L\pm}^{\bar{X}\bar{M}}$ for the $\bar{X}\bar{M}$ line cut have a nearly linear energy dispersion with a slope $\approx v_x/2$. Similar to the case of $\bar{\Gamma}\bar{X}$ scattering, we can distinguish these three features by examining their intercepts to $q = 0$ and $2\pi/d$. Linear extrapolation of the high energy features of $q_{H-}^{\bar{X}\bar{M}}$, $q_{1L-}^{\bar{X}\bar{M}}$, and $q_{2L-}^{\bar{X}\bar{M}}$ to $q = 0$ or $2\pi/d$ gives $m$ (negative value), 0, and $-m$ (positive value), respectively. The negative intercepts of the $q_{H-}$ and $q'_{H-}$ peaks rule out $q_{1L-}^{\bar{X}\bar{M}}$ and $q_{2L-}^{\bar{X}\bar{M}}$. Therefore, we can attribute the experimental $q_{H-}$ and $q'_{H-}$ peaks to the $q_{H-}^{\bar{X}\bar{M}}$ critical spanning vector for this case of $\bar{X}\bar{M}$ scattering and get the Fermi velocity along the $\bar{X}\bar{M}$ direction $v_x = 1.51 \pm 0.08$ eV·Å from the slope and $m = -71 \pm 9$ meV from the intercept. We can then get the Fermi velocity along the $\bar{\Gamma}\bar{X}$ direction $v_y$ as a function of the Lifshitz transition energy $\delta$ using $v_y = \pm\sqrt{m^2 + \delta^2}/\bar{\Lambda}_{1,2}$, where $m = -71 \pm 9$ meV and $\bar{\Lambda}_{1,2} = 0.180 \pm 0.003$ Å$^{-1}$.[30] Table 1 shows $v_y$ and $E_{H-}$ as a function of $\delta$. To determine a range of $\delta$ that is consistent with our data, we plot the energy dispersion of the critical spanning vector $q_{H-}^{\bar{X}\bar{M}}$ with different choices of $\delta$ as dashed lines in Fig. 8(b). The plots follow the data for $\delta = 20$ meV and 40 meV, but start to deviate from the data for $\delta \geq 60$ meV. Therefore, we determine the range of the Lifshitz transition energy $\delta$ to be $\approx 0$ meV to 60 meV. With both $v_x$ and $v_y$ deduced from our data, we plot the energy dispersions of the critical spanning vectors for both $\bar{\Gamma}\bar{X}$ (solid lines) and $\bar{X}\bar{M}$ (dashed lines) line cuts on top of the experimental data for the range of $q > 0$ (right hand portion) in Fig. 9(e), with $\delta = 30$ meV, $m = -71$ meV, $v_x = 1.51$ eV·Å, and $v_y = 0.428$ eV·Å. Now the $q$ location of the $q_3$ feature at the zero energy crossing point is consistent with our data and the $q_{H-}$ and $q'_{H-}$ peaks are well described by the energy dispersion of the $q_{H-}^{\overline{XM}}$ critical spanning vector (dashed green lines). The domination of the $q_{H\pm}^{\overline{XM}}$ features over the $q_{H\pm}^{\overline{\Gamma X}}$ features can be expected because of the better nesting of the Fermi surface along the $\overline{XM}$ direction. However, we did not observe any strong energy dependence of the other features such as $q_1$ and $q_3$ as suggested by the critical spanning vectors. We also note the Fermi velocities deduced from our data for this case is much smaller than those obtained from ARPES measurements[7,8].

In summary, both cases of scattering along the $\overline{\Gamma X}$ and $\overline{XM}$ line cuts can describe the observed $q_{H-}$ and $q'_{H-}$ features. However, neither explains our data completely. The case for the $\overline{\Gamma X}$ scattering would indicate that the $q_3$ features are closer to the Bragg peaks than what was observed in our quasiparticle interference patterns. The case for the $\overline{XM}$ scattering would indicate smaller Fermi velocities than those reported in the literature[7,8,31], which give rise to strong energy dependent $q$ features and interconnected patterns at high energies [see Figs. 9(e) and 10] that were not observed in the data. Therefore it is likely that the actual scattering interference phenomena observed in STS measurements has contributions from both of these extreme cases. In the next section, we will simulate quasiparticle interference patterns to shine light on these two possible interpretations.

**C. Surface states spin texture**

To probe the possible spin textures at the surface, we have also carried out calculations of the joint density of states. The joint density of states is closely related to quasiparticle interference patterns because scattering $q$ wave vectors that connect regions of high density of states on the contours of constant energy contribute to a large degree in the joint density of states maps. The joint density of states is computed by taking the autoconvolution of the initial and final scattering states[29]

$$JDOS(\mathbf{q}, E) = \int \rho(\mathbf{k}, E)\rho(\mathbf{k}+\mathbf{q}, E)d^2\mathbf{k}, \qquad (2)$$

where $\rho(\mathbf{k}, E)$ and $\rho(\mathbf{k}+\mathbf{q}, E)$ are the initial and final density of states. The density of states for Eq. (2) is obtained as a constant around the contours of constant energy derived from the energy-momentum dispersions of the surface bands described by Equation (1). To compare the calculation with our experimental data, we show the computed joint density of states in Figs. 10(a) and 10(c) for different energies away from the zero energy crossing point. For the case of scattering along the $\overline{\Gamma X}$ line cut [Fig. 10(a)], we cannot obtain $v_x$ directly from our data. But given that our experimental values of $v_y = 1.51 \pm 0.08$ eV·Å and $\sqrt{m^2 + \delta^2} = 71 \pm 9$ meV are close to $v_y = 1.30$ eV·Å and $\sqrt{m^2 + \delta^2} = 75$ meV in Ref. [31], it is reasonable to choose $m = -70$ meV, $\delta = 26$ meV, and $v_x = 2.40$ eV·Å from Ref. [31] with an experimental value of $v_y = 1.51$ eV·Å for the joint density of states calculation to have a qualitative comparison with the experimental quasiparticle interference patterns. As for the case of scattering along the $\overline{XM}$ line cut [Fig. 10(c)], we choose $m = -71$ meV, $\delta = 30$ meV, $v_x = 1.51$ eV·Å, and $v_y = 0.428$ eV·Å obtained from our data. The white dashed boxes indicate the first scattering zones with a size of $4\pi/d \times 4\pi/d$. The calculation suggests a rich structure in the quasiparticle interference patterns. When the energy is decreased from the zero energy crossing point, the general trend is that the disk size for all the $q$ wave vectors becomes larger as the surface state Fermi pockets grow larger. The calculated patterns using the smaller Fermi velocities obtained from the $\overline{XM}$ scattering [Fig. 10(c)] generally have a larger disk size compared to those obtained with larger Fermi velocities in Fig. 10 (a). Fig. 10(c) also shows that different $q$ features such as $q_1$ and $q_3$ become interconnected to each other for energies far away from the zero energy crossing point.

To understand the impact of the spin-momentum locked surface states on the scattering process, we also take the spin texture into account and compute the spin selective joint density of states following Ref. [29],

$$SJDOS(\mathbf{q}) = \int \rho(\mathbf{k})T(\mathbf{q},\mathbf{k})\rho(\mathbf{k}+\mathbf{q})d^2\mathbf{k}, \qquad (3)$$

where $T(\mathbf{q},\mathbf{k})$ is the spin-dependent scattering matrix element. We model the spin texture as simple artificial momentum-locked spins that are tangential to the contours of constant energy. Due to the spin-momentum locking mechanism, the scattering process is suppressed for the unaligned spins and forbidden for oppositely aligned spins. Thus, this spin-dependent scattering causes the quasiparticle interference pattern to differ from the one obtained from the joint density of states without considering spin directions. As shown in Figs. 10(b) and 10(d), the computed spin selective joint density of states with the same two sets of parameters obviously contrast the corresponding ones obtained from the joint density of states. The spin selective joint density of states disk size for $q_2$ and $q_3$ is smaller over the entire energy range compared to that of the joint density of states, which indicates reduced scattering at the surface. Furthermore, the changed shape of the $q$ vectors as compared to that of the joint density of states also suggests reduced intra- and inter-cone scattering. A direct comparison between the experimental and computed quasiparticle interference patterns suggests the choice of larger Fermi velocities for the surface bands (the $\overline{\Gamma}\overline{X}$ scattering case) agrees better with our experimental quasiparticle interference patterns though the observed $\overline{\Lambda_1}\overline{X}$ value is larger than expected based on the model in Ref. [31]. This discrepancy may be related to the warping in the Fermi surface of the surface bands and/or the limitation of the model. The small disk size of $q_2$ and $q_3$ in our experimental quasiparticle interference patterns also suggests reduced scattering at the surface, possibly due to the spin-momentum locked topological surface states. However, the relatively weak scattering from defects, smearing of features due to surface disorder, and high doping concentration in our sample preclude a definitive confirmation of the spin texture of the surface states. Further measurements on samples with very low doping concentration are necessary to confirm the spin texture of the SnTe topological surface states.

### IV. Conclusions

In conclusion, we have synthesized single crystalline SnTe nanoplates on graphitized 6H-SiC substrates by MBE and carried out *in-situ* STM measurements on the (001) surface states. Our observation of the quasiparticle interference patterns is consistent with scattering among the four Fermi pockets of the surface states in the first surface Brillouin zone. The energy dispersion of the quasiparticle interference intensity shows two high energy features related to the crossing point beyond the Lifshitz transition when the two neighboring low energy surface bands near the $\overline{X}$ point merge. We have presented two possible interpretations for the two high energy features due to different scattering vectors. A comparison between the experimental and computed quasiparticle interference patterns seems to suggest the case of $\overline{\Gamma}\overline{X}$ scattering agrees better with our data as well as possible spin texture of the surface states. This work demonstrates that SnTe nanoplates can provide a model system for studying topological crystalline insulator surface states and exploring potential device applications.

**Acknowledgement**

We thank Liang Fu, Mark Stiles, and Guru Khalsa for valuable comments and discussions. We are also grateful to Takafumi Sato and Yoichi Ando for sharing with us the SnTe (001) surface state parameters estimated from their ARPES measurements. D. Z., J. H., and T. Z. acknowledge support under the Cooperative Research Agreement between the University of Maryland and the National Institute of Standards and Technology Center for Nanoscale Science and Technology, Grant No. 70NANB10H193, through the University of Maryland. H. B. and Y. K. are partly supported by Korea Research Foundation through Grant No. KRF-2010-00349.

**Table Captions**

**Table 1** The Fermi velocity along the $\bar{\Gamma}\bar{X}$ direction $v_y$ and the high energy crossing point energy $E_{H^-}$ as a function of the Lifshitz transition energy $\delta$.

**Figure Captions**

**Fig. 1** (Color online). Synthesis and characterization of SnTe nanoplates. (a) Schematic of the SnTe rock salt structure. (b) RHEED pattern from SnTe nanoplates during MBE growth. (c) AFM image of as grown SnTe nanoplates. (d) XRD from SnTe nanoplates confirms single crystal nature showing only {001} reflections. The asterisks indicate the {0001} reflections of the SiC substrate.

**Fig. 2** (Color online). (a) A 100 nm × 100 nm STM topographic image showing terraces with step height of one half unit cell (≈ 0.32 nm). The color scale covers a height range of 1.15 nm. (b) A typical 55 nm × 55 nm STM topographic image at the (001)-terminated surface of a SnTe nanoplate, tunneling setpoint: $V_{\text{bias}} = 0.8$ V and $I = 100$ pA. The color scale covers a height range

of 98 pm. (c) Schematic SnTe (001) surface state band structure. The first surface Brillouin zone is indicated by the shaded square. (d) Single point $dI/dV$ spectrum, tunneling setpoint: $V_{bias} = 0.7$ V and $I = 150$ pA. Inset shows a simplified schematic band diagram of the two low energy surface bands located symmetrically about the $\bar{X}$ point along $\bar{\Gamma}\bar{X}$ direction. $E_0$ denotes the zero energy crossing point and $E_{H\pm}$ denotes the high energy crossing points beyond the Lifshitz transition. Blue and red lines indicate opposite spin orientations.

**Fig. 3** (Color online). Atomic defects on the SnTe (001) surface. (a) and (c) Experimental high-resolution topographic images of atomic defects on the SnTe (001) surface at $I = 70$ pA and different sample biases. The color scale range is 65 pm for $V_{bias} = 1.0$ V, 60 pm for $V_{bias} = 0.8$ V, 65 pm for $V_{bias} = 0.6$ V, 70 pm for $V_{bias} = 0.4$ V, 40 pm for $V_{bias} = -0.8$ V, 50 pm for $V_{bias} = -0.6$ V, 60 pm for $V_{bias} = -0.4$ V, and 55 pm for $V_{bias} = -0.2$ V. The Sn vacancies are indicated by red arrows and other types of defects are indicated by green arrows in image $V_{bias} = 1.0$ V. The dashed lines indicate the Te atom site positions. (b) and (d) DFT simulated topographic images with Sn vacancies at different sample biases. The color scale range is 162 pm and the dashed lines indicate the Te atom site positions.

**Fig. 4** (Color online). $dI/dV$ spatial maps of a 47.5 nm × 47.5 nm area at different energies with respect to the zero energy crossing point $E_0$, which is 350 meV above the Fermi level.

**Fig. 5** (Color online). Quasiparticle interference at the SnTe (001) surface. (a) Schematic of SnTe (001) surface band structure at energies below the Lifshitz transition in the first surface Brillouin zone. $q_1$ represents intra-cone scattering; $q_2$, $q_3$, and $q_4$ represent inter-cone scattering. (b) Schematic contours of constant energy near the zero energy crossing point in the first surface Brillouin zone. $q$ vectors are indicated by the color coded arrows. Due to periodic Brillouin zone boundary condition, $q_3$ is equivalent to $q'_3$, the inter-cone scattering across the first surface Brillouin zone boundary. (c) Schematic quasiparticle interference pattern enclosed by a box with size of $4\pi/d \times 4\pi/d$ in $q$ space.

**Fig. 6** (Color online). Experimental quasiparticle interference patterns (unfiltered) at $E_0 + 100$ meV (a), $E_0 + 50$ meV (b), $E_0$ (c), $E_0 - 50$ meV (d), $E_0 - 100$ meV (e), $E_0 - 150$ meV (f), $E_0 - 200$ meV (g), and $E_0 - 250$ meV (h). $E_0 = 350$ meV above the Fermi level.

**Fig. 7** (Color online). (a) Energy dispersion of quasiparticle interference intensity along the $\overline{\Gamma}\overline{M}$ ($q_2$) direction. (b) Energy dispersion of quasiparticle interference intensity along the $\overline{\Gamma}\overline{X}$ ($q_3$) direction.

**Fig. 8** (Color online). (a) Slices of the quasiparticle interference intensity extracted from Fig. 7(b) at energies from $E_0 - 300$ meV to $E_0 - 200$ meV. The arrows indicate the Bragg peak, $q_3$, $q_{H-}$ and $q'_{H-}$ features. The inset plots the sum of $q_{H-}$ and $q'_{H-}$ vs. energy, and the red dashed line indicates the position of the Bragg peak. (b) $E$ vs. $q$ data extracted from the energy dispersion of $q_{H-}$ peak position. The one standard deviation uncertainty in the $q$ wave vector obtained from the peak fitting is less than the symbol size in the plot. The red line is a linear fit which yields $v_{F\text{-}H} = 1.51 \pm 0.08$ eV·Å.[32] The dashed lines are the energy dispersion of the critical spanning vector $q_{H-}^{\overline{X}\overline{M}}$ with different choices of $\delta$. (c) $E$ vs. $q$ data extracted from the energy dispersion of $q'_{H-}$ peak position. The red line is a linear fit which yields $v_{F\text{-}H} = 1.28 \pm 0.28$ eV·Å.[32]

**Fig. 9** (Color online). (a) Schematic contours of constant energy of the surface bands at energy above the high energy crossing point. The box indicates the first surface Brillouin zone. (b) Schematic band diagram showing critical spanning vectors along the $\overline{\Gamma}\overline{X}$ line cut [dashed blue line in panel (a)] in $k$ space. Red arrows represent $q_1^{\overline{\Gamma}\overline{X}}$ and green arrows represent intra-cone scatterings of $CONE_{H+}$ and $CONE_{H-}$. $q_{H\pm}^{\overline{\Gamma}\overline{X}}$ at $E_{H-} < E < E_{H+}$ is a subset of $q_3^{\overline{\Gamma}\overline{X}}$ defined earlier in Fig. 5. (c) Schematic band diagram showing critical spanning vectors along the $\overline{X}\overline{M}$ line cut [dashed red line in panel (a)] in $k$ space. Possible scatterings are indicated by the color coded arrows. (d) Energy dispersion of the critical spanning vectors. Solid and dashed lines represent the critical spanning vectors along the line cuts of $\overline{\Gamma}\overline{X}$ and $\overline{X}\overline{M}$ directions, respectively. The origins of the critical spanning vectors are indicated by the arrows. (e) Energy dispersion of the critical spanning vectors superimposed on the experimental data along the $\overline{\Gamma}\overline{X}$ direction. Parameters for $q < 0$:

$v_y = 1.51$ eV·Å and $\sqrt{m^2 + \delta^2} = 71$ meV. Parameters for $q > 0$: $\delta = 30$ meV, $m = -71$ meV, $v_x = 1.51$ eV·Å, and $v_y = 0.428$ eV·Å. Solid and dashed lines indicate energy dispersion of the critical spanning vectors along the $\overline{\Gamma}\overline{X}$ and $\overline{X}\overline{M}$ line cuts, respectively. Solid violet lines: $q_1^{\overline{\Gamma}\overline{X}}$. Solid green lines: $q_{H\pm}^{\overline{\Gamma}\overline{X}}$. Dashed green lines: $q_{H\pm}^{\overline{X}\overline{M}}$. Dashed violet lines: $q_{1L\pm}^{\overline{X}\overline{M}}$. Dashed blue lines: $q_{2L\pm}^{\overline{X}\overline{M}}$ and $q_{3L\pm}^{\overline{X}\overline{M}}$.

**Fig. 10** (Color online). Computed quasiparticle interference patterns. (a) Computed joint density of states (JDOS) without taking spin into account at different energies with respect to the zero energy crossing point. Parameters: $m = -70$ meV, $\delta = 26$ meV, $v_x = 2.40$ eV·Å, and $v_y = 1.51$ eV·Å. (b) Computed spin selective joint density of states (SJDOS) with the spin texture of the surface states taken into account at different energies with respect to the zero energy crossing point. Parameter choice is the same as panel (a). (c) Computed joint density of states without taking spin into account at different energies with respect to the zero energy crossing point. Parameters: $m = -71$ meV, $\delta = 30$ meV, $v_x = 1.51$ eV·Å, and $v_y = 0.428$ eV·Å. (d) Computed spin selective joint density of states with the spin texture of the surface states taken into account at different energies with respect to the zero energy crossing point. Parameters are the same as in panel (c).

**Table 1** The Fermi velocity along the $\overline{\Gamma}\overline{X}$ direction $v_y$ and the high energy crossing point energy $E_{H\text{-}}$ as a function of the Lifshitz transition energy $\delta$.

| $\delta$ (meV) | $v_y$ (eV·Å) | $E_{H\text{-}}$ (meV) |
|---|---|---|
| 10 | 0.398 | -72 |
| 20 | 0.410 | -74 |
| 30 | 0.428 | -77 |
| 40 | 0.453 | -81 |
| 50 | 0.482 | -87 |
| 60 | 0.516 | -93 |
| 80 | 0.594 | -107 |
| 100 | 0.681 | -123 |

Table 1

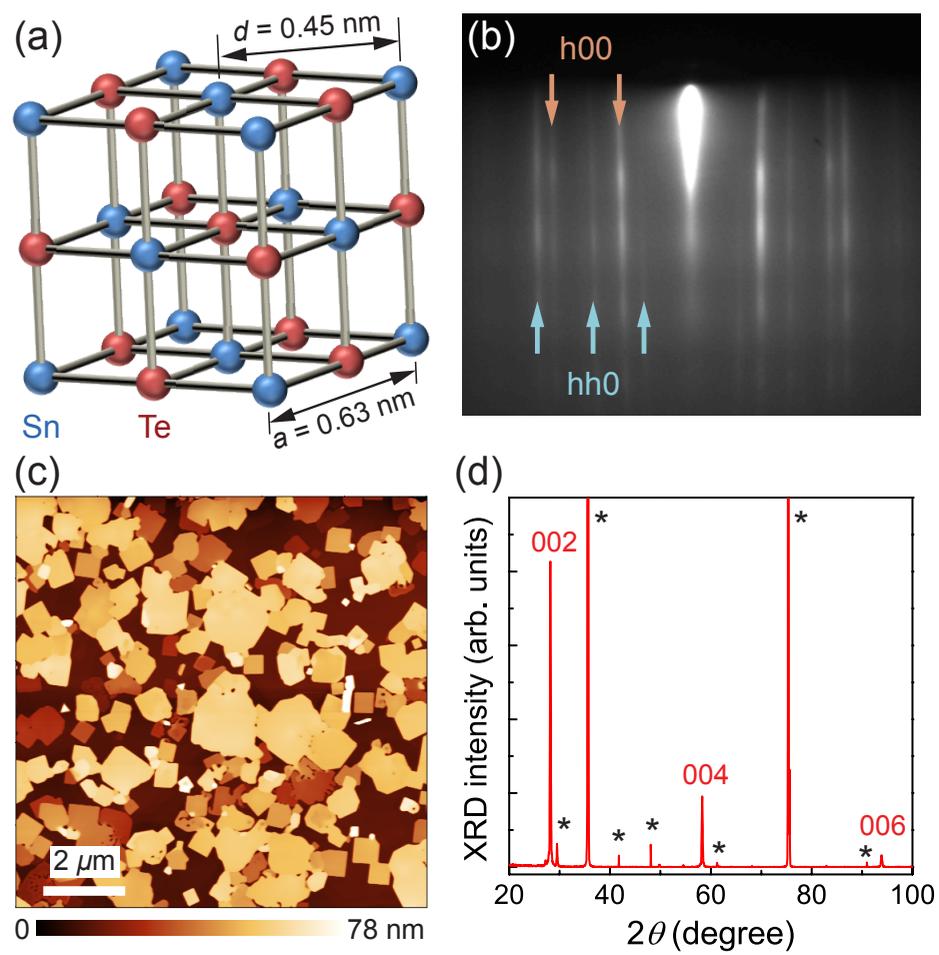

FIG. 1 (Color online). Synthesis and characterization of SnTe nanoplates. (a) Schematic of the SnTe rock salt structure. (b) RHEED pattern from SnTe nanoplates during MBE growth. (c) AFM image of as grown SnTe nanoplates. (d) XRD from SnTe nanoplates confirms single crystal nature showing only {001} reflections. The asterisks indicate the {0001} reflections of the SiC substrate.

*Print one column*

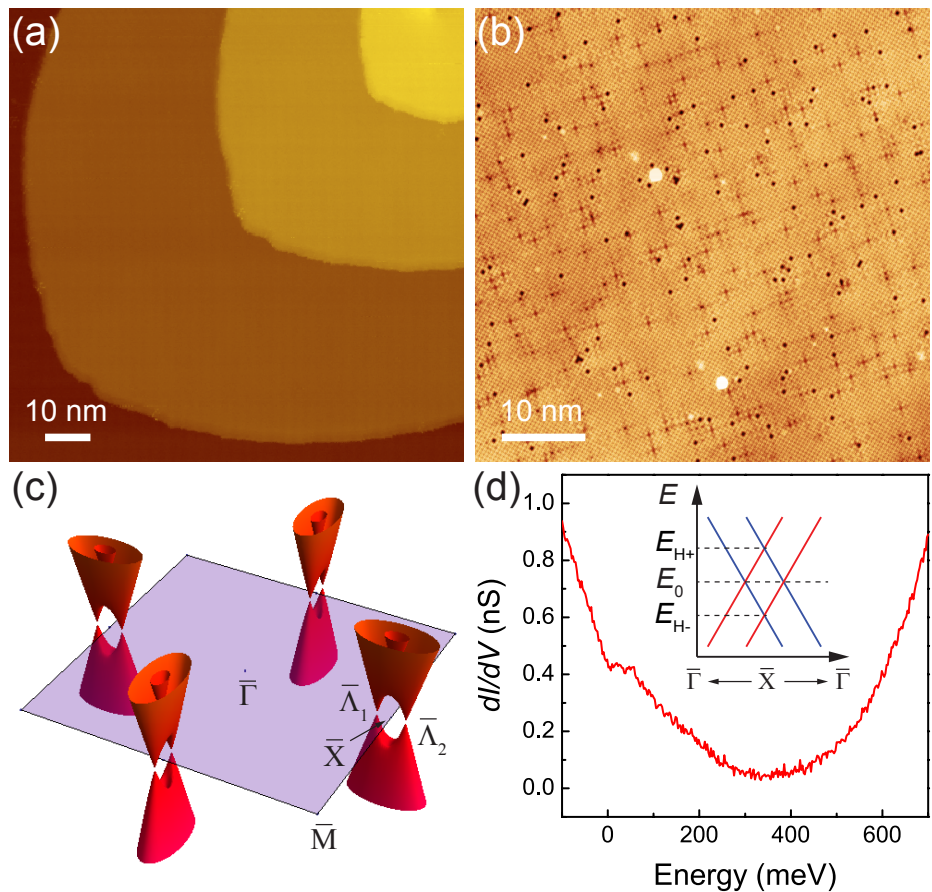

FIG. 2 (Color online). (a) A 100 nm × 100 nm STM topographic image showing terraces with step height of one half unit cell (≈ 0.32 nm). The color scale covers a height range of 1.15 nm. (b) A typical 55 nm × 55 nm STM topographic image at the (001)-terminated surface of a SnTe nanoplate, tunneling setpoint: $V_{bias}$ = 0.8 V and $I$ = 100 pA. The color scale covers a height range of 98 pm. (c) Schematic SnTe (001) surface state band structure. The first surface Brillouin zone is indicated by the shaded square. (d) Single point $dI/dV$ spectrum, tunneling setpoint: $V_{bias}$ = 0.7 V and $I$ = 150 pA. Inset shows a simplified schematic band diagram of the two low energy surface bands located symmetrically about the $\bar{X}$ point along $\bar{\Gamma}\bar{X}$ direction. $E_0$ denotes the zero energy crossing point and $E_{H\pm}$ denotes the high energy crossing points beyond the Lifshitz transition. Blue and red lines indicate opposite spin orientations.

*Print one column*

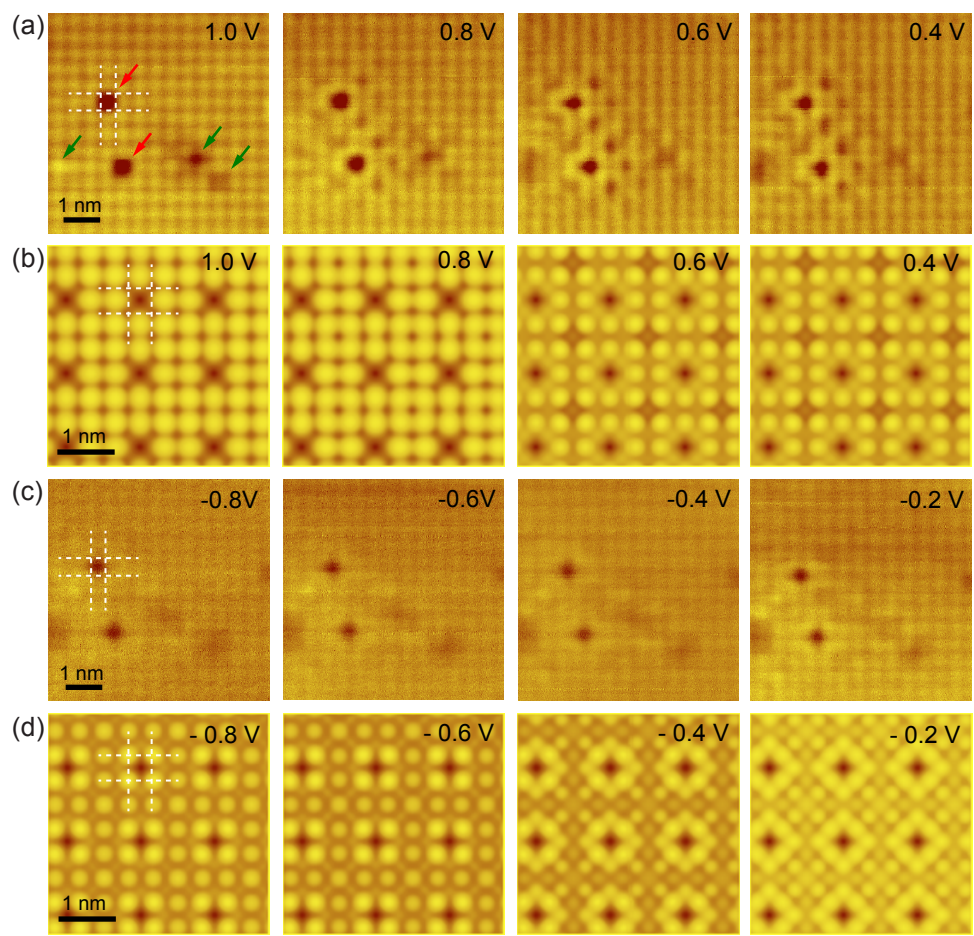

FIG. 3 (Color online). Atomic defects on the SnTe (001) surface. (a) and (c) Experimental high-resolution topographic images of atomic defects on the SnTe (001) surface at $I = 70$ pA and different sample biases. The color scale range is 65 pm for $V_{bias} = 1.0$ V, 60 pm for $V_{bias} = 0.8$ V, 65 pm for $V_{bias} = 0.6$ V, 70 pm for $V_{bias} = 0.4$ V, 40 pm for $V_{bias} = -0.8$ V, 50 pm for $V_{bias} = -0.6$ V, 60 pm for $V_{bias} = -0.4$ V, and 55 pm for $V_{bias} = -0.2$ V. The Sn vacancies are indicated by red arrows and other types of defects are indicated by green arrows in image $V_{bias} = 1.0$ V. The dashed lines indicate the Te atom site positions. (b) and (d) DFT simulated topographic images with Sn vacancies at different sample biases. The color scale range is 162 pm and the dashed lines indicate the Te atom site positions.

*Print 1.5 columns*

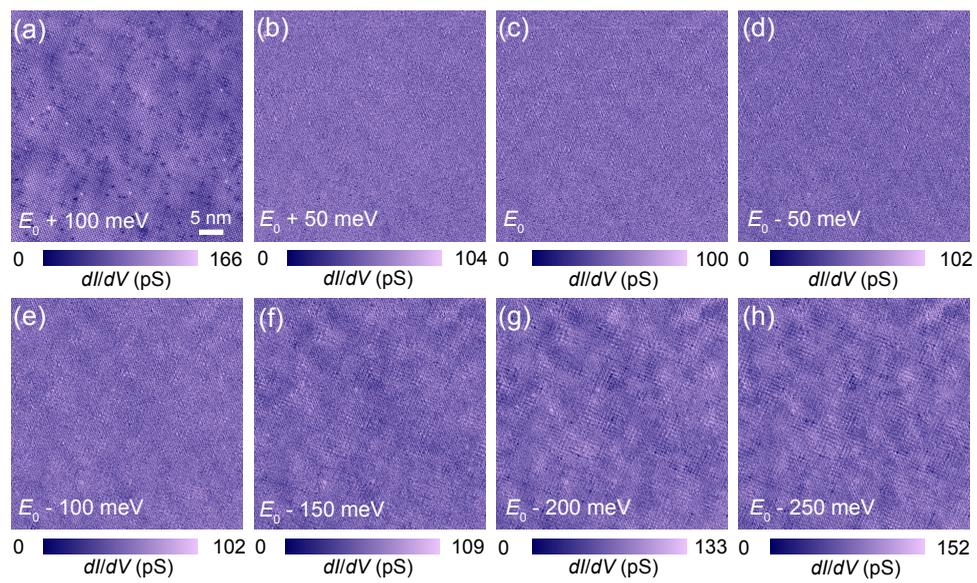

FIG. 4 (Color online). *dI/dV* spatial maps of a 47.5 nm × 47.5 nm area at different energies with respect to the zero energy crossing point $E_0$, which is 350 meV above the Fermi level.

*Print 1.5 columns*

FIG. 5 (Color online). Quasiparticle interference at the SnTe (001) surface. (a) Schematic of SnTe (001) surface band structure at energies below the Lifshitz transition in the first surface Brillouin zone. $q_1$ represents intra-cone scattering; $q_2$, $q_3$, and $q_4$ represent inter-cone scattering. (b) Schematic contours of constant energy near the zero energy crossing point in the first surface Brillouin zone. $q$ vectors are indicated by the color coded arrows. Due to periodic Brillouin zone boundary condition, $q_3$ is equivalent to $q'_3$, the inter-cone scattering across the first surface Brillouin zone boundary. (c) Schematic quasiparticle interference pattern enclosed by a box with size of $4\pi/d \times 4\pi/d$ in $q$ space.

*Print one column*

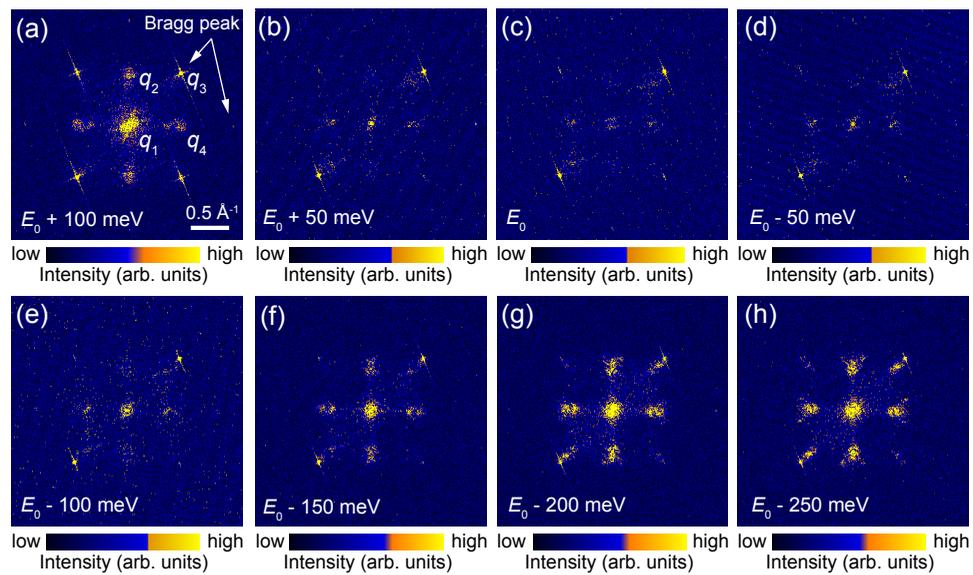

FIG. 6 (Color online). Experimental quasiparticle interference patterns (unfiltered) at $E_0 + 100$ meV (a), $E_0 + 50$ meV (b), $E_0$ (c), $E_0 - 50$ meV (d), $E_0 - 100$ meV (e), $E_0 - 150$ meV (f), $E_0 - 200$ meV (g), and $E_0 - 250$ meV (h). $E_0 = 350$ meV above the Fermi level.

*Print 1.5 columns*

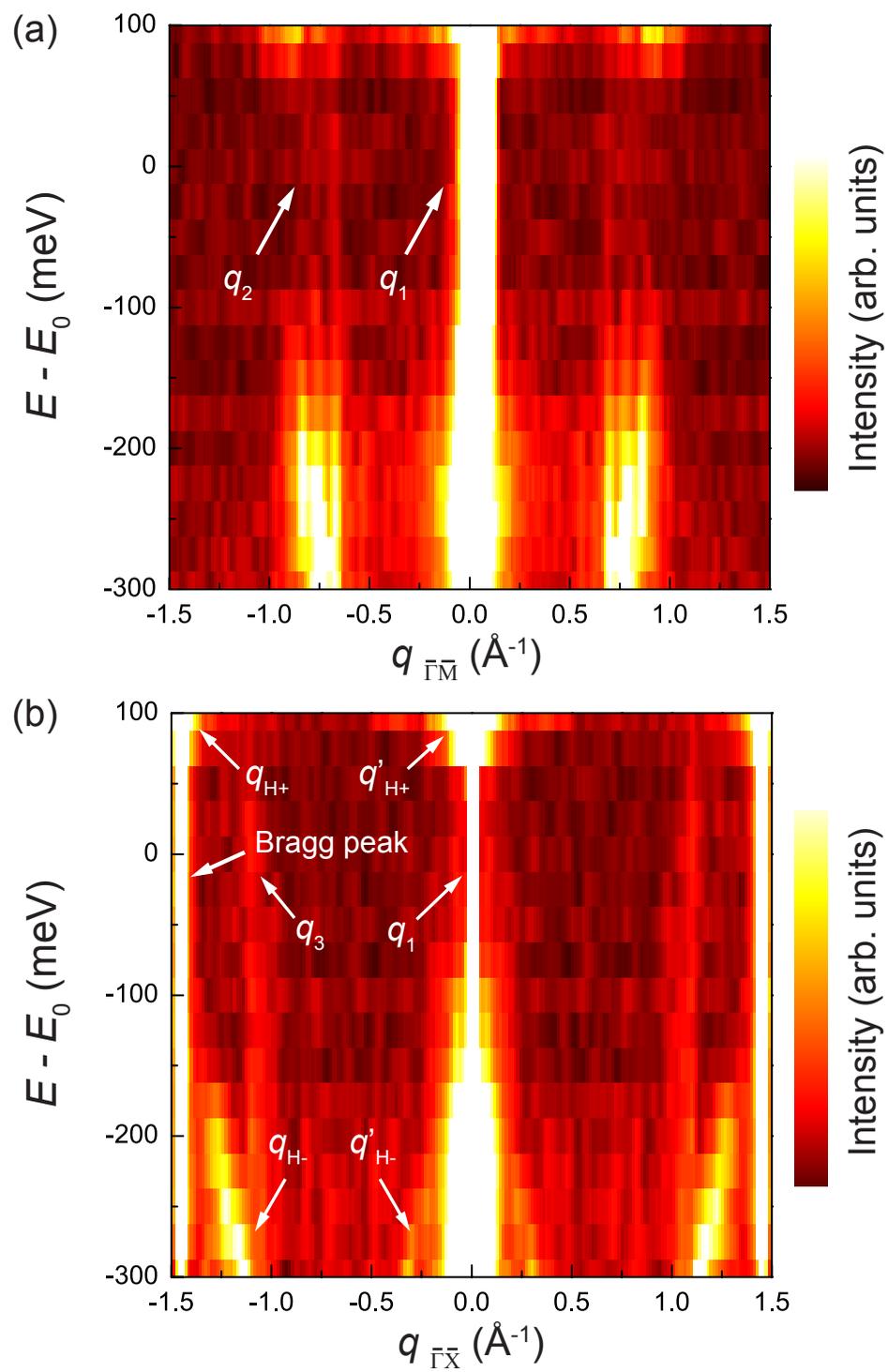

FIG. 7 (Color online). (a) Energy dispersion of quasiparticle interference intensity along the $\bar{\Gamma}\bar{M}$ ($q_2$) direction. (b) Energy dispersion of quasiparticle interference intensity along the $\bar{\Gamma}\bar{X}$ ($q_3$) direction.

*Print one column*

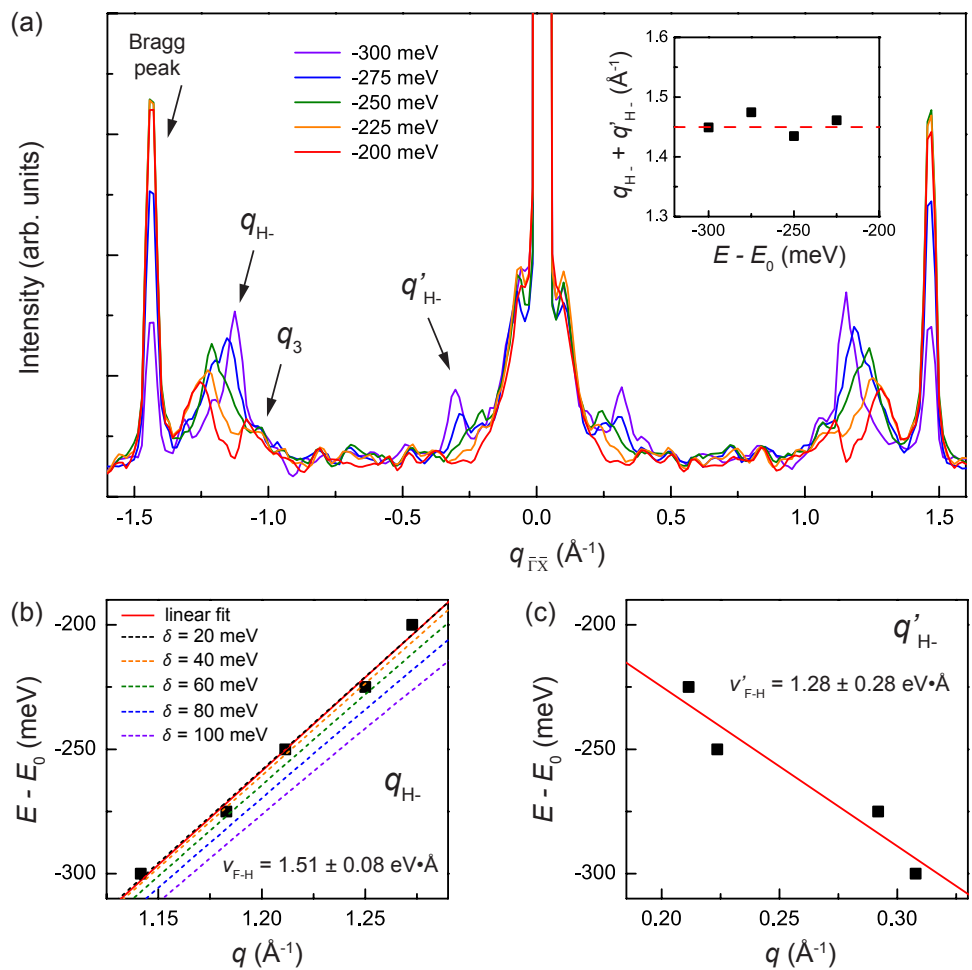

FIG. 8 (Color online). (a) Slices of the quasiparticle interference intensity extracted from Fig. 7(b) at energies from $E_0 - 300$ meV to $E_0 - 200$ meV. The arrows indicate the Bragg peak, $q_3$, $q_{H-}$ and $q'_{H-}$ features. The inset plots the sum of $q_{H-}$ and $q'_{H-}$ vs. energy, and the red dashed line indicates the position of the Bragg peak. (b) $E$ vs. $q$ data extracted from the energy dispersion of $q_{H-}$ peak position. The one standard deviation uncertainty in the $q$ wave vector obtained from the peak fitting is less than the symbol size in the plot. The red line is a linear fit which yields $v_{F-H} = 1.51 \pm 0.08$ eV·Å.[32] The dashed lines are the energy dispersion of the critical spanning vector $q_{H-}^{\bar{X}\bar{M}}$ with different choices of $\delta$. (c) $E$ vs. $q$ data extracted from the energy dispersion of $q'_{H-}$ peak position. The red line is a linear fit which yields $v'_{F-H} = 1.28 \pm 0.28$ eV·Å.[32]

*Print 1.5 columns*

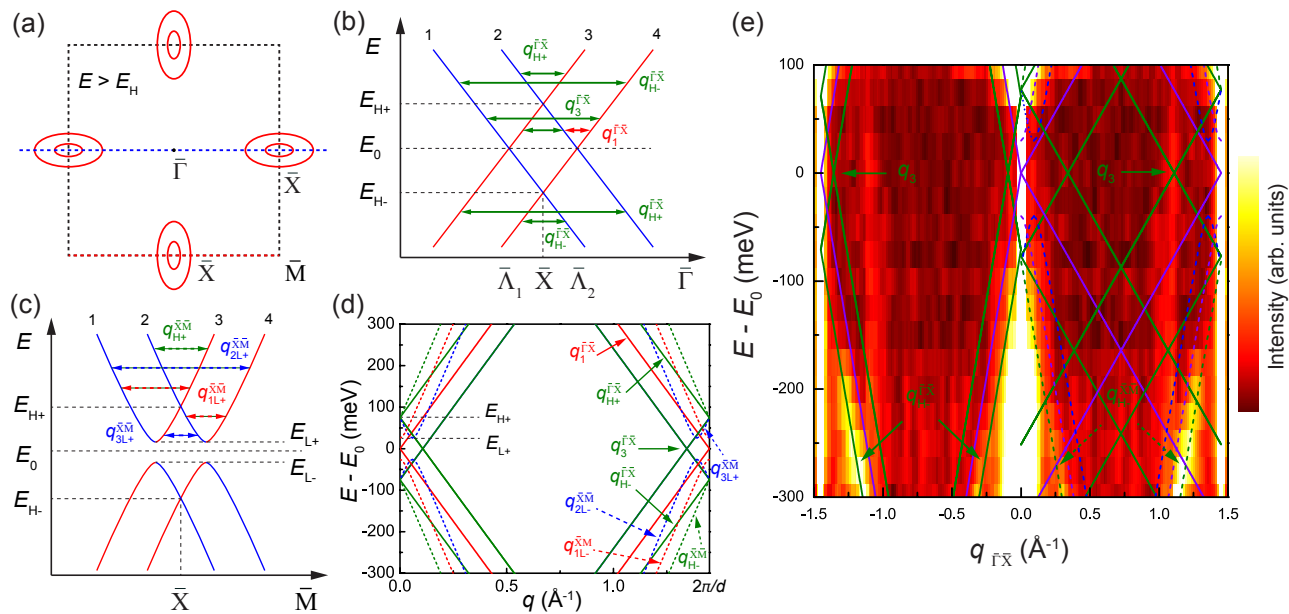

FIG. 9 (Color online). (a) Schematic contours of constant energy of the surface bands at energy above the high energy crossing point. The box indicates the first surface Brillouin zone. (b) Schematic band diagram showing critical spanning vectors along the $\bar{\Gamma}\bar{X}$ line cut [dashed blue line in panel (a)] in $k$ space. Red arrows represent $q_1^{\bar{\Gamma}\bar{X}}$ and green arrows represent intra-cone scatterings of $CONE_{H+}$ and $CONE_{H-}$. $q_{H\pm}^{\bar{\Gamma}\bar{X}}$ at $E_{H-} < E < E_{H+}$ is a subset of $q_3^{\bar{\Gamma}\bar{X}}$ defined earlier in Fig. 5. (c) Schematic band diagram showing critical spanning vectors along the $\bar{X}\bar{M}$ line cut [dashed red line in panel (a)] in $k$ space. Possible scatterings are indicated by the color coded arrows. (d) Energy dispersion of the critical spanning vectors. Solid and dashed lines represent the critical spanning vectors along the line cuts of $\bar{\Gamma}\bar{X}$ and $\bar{X}\bar{M}$ directions, respectively. The origins of the critical spanning vectors are indicated by the arrows. (e) Energy dispersion of the critical spanning vectors superimposed on the experimental data along the $\bar{\Gamma}\bar{X}$ direction. Parameters for $q < 0$: $v_y$ = 1.51 eV·Å and $\sqrt{m^2 + \delta^2}$ = 71 meV. Parameters for $q > 0$: $\delta$ = 30 meV, $m$ = -71 meV, $v_x$ = 1.51 eV·Å, and $v_y$ = 0.428 eV·Å. Solid and dashed lines indicate energy dispersion of the critical spanning vectors along the $\bar{\Gamma}\bar{X}$ and $\bar{X}\bar{M}$ line cuts, respectively. Solid violet lines: $q_1^{\bar{\Gamma}\bar{X}}$. Solid green lines: $q_{H\pm}^{\bar{\Gamma}\bar{X}}$. Dashed green lines: $q_{H\pm}^{\bar{X}\bar{M}}$. Dashed violet lines: $q_{1L\pm}^{\bar{X}\bar{M}}$. Dashed blue lines: $q_{2L\pm}^{\bar{X}\bar{M}}$ and $q_{3L\pm}^{\bar{X}\bar{M}}$.

*Print two columns*

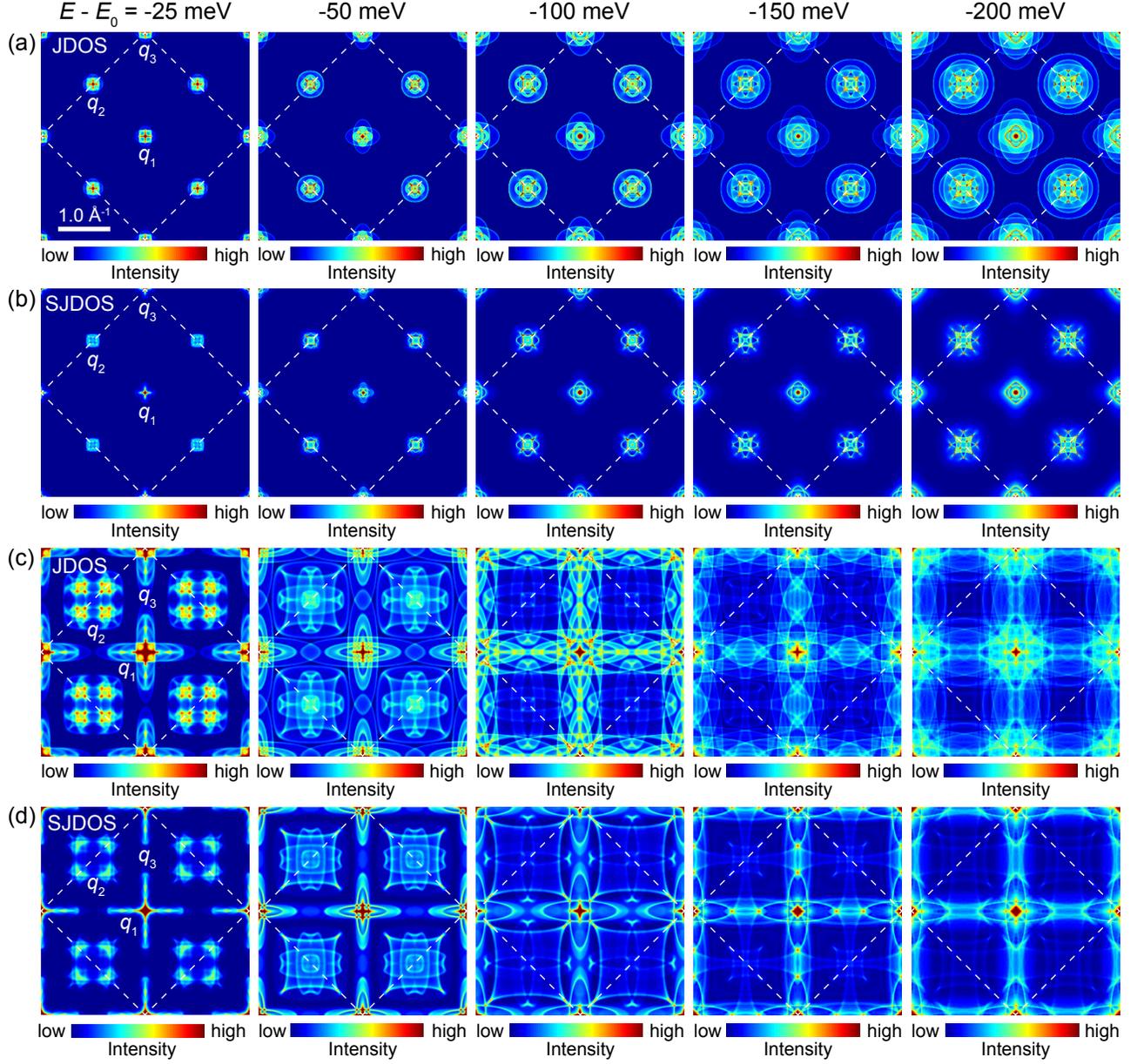

FIG. 10 (Color online). Computed quasiparticle interference patterns. (a) Computed joint density of states (JDOS) without taking spin into account at different energies with respect to the zero energy crossing point. Parameters: $m = -70$ meV, $\delta = 26$ meV, $v_x = 2.40$ eV·Å, and $v_y = 1.51$ eV·Å. (b) Computed spin selective joint density of states (SJDOS) with the spin texture of the surface states taken into account at different energies with respect to the zero energy crossing point. Parameter choice is the same as panel (a). (c) Computed joint density of states without taking spin into account at different energies with respect to the zero energy crossing point. Parameters: $m = -71$ meV, $\delta = 30$ meV, $v_x = 1.51$ eV·Å, and $v_y = 0.428$ eV·Å. (d) Computed spin selective joint density of states with the spin texture of the surface states taken into account at different energies with respect to the zero energy crossing point. Parameters are the same as in panel (c).

*Print two columns*